\documentclass[a4paper]{jpconf}
\usepackage{graphicx}
\begin{document}
\title{The discovery and characterisation of binary central stars in planetary nebulae}

\author{David Jones$^{1,2}$}

\address{$^1$ Instituto de Astrof\'isica de Canarias, E-38205 La Laguna, Tenerife, Spain\\
$^2$ Departamento de Astrof\'isica, Universidad de La Laguna, E-38206 La Laguna, Tenerife, Spain
}

\ead{djones@iac.es}

\begin{abstract}
Close binary central stars of planetary nebulae are key in constraining the poorly-understood common-envelope phase of evolution, which in turn is critical in understanding the formation of a wide-range of astrophysical phenomena (including cataclysmic variables, low-mass X-ray binaries and supernovae type Ia). Here, I present the results of our on-going, targeted search for close-binaries in planetary nebulae which has led to the discovery of more than 10 new central binaries in just the last few years (almost the same as the total discovered during the 1980s and 1990s together). This success has been rooted in the targeted selection of objects for study, based on morphological features deemed typical of binarity, as well as novel observing strategies (including the employment of narrow-band filters for photometry to minimise nebular contamination), both of which are discussed. These new discoveries coupled with the painstaking characterisation of both newly discovered systems and those from the literature mean that we are now in a position to begin to probe the poorly understood common-envelope phase.
\end{abstract}

\section{Introduction}

While central star binarity has long been considered a key factor in the formation of aspherical morphologies in planetary nebulae (PNe), it is only recently that it has been shown observationally that a close-binary pathway is responsible for a significant fraction of PNe \cite{miszalski09a} \footnote{Note that in these proceedings, I will only cover the close-binary pathway.  However, long-term radial velocity surveys have also produced very promising results in the search for long period binaries \cite{vanwinckel14}.}.  This lag was, in part, due to the painstakingly difficult nature of the search for binary central stars - with photometric monitoring generally limited to the brightest central stars and fainter nebulae for ease of photometry with modestly sized telescopes.  This all changed with the advent of the OGLE survey, an i-band monitoring survey of the Galactic bulge primarily in search of gravitational lensing events, but which also resulted in the discovery of more than 10 new close-binary central stars - approximately doubling the sample known at the time \cite{miszalski09a} (see figure 1 of \cite{jones15b}).  With the resulting sample, it was possible to identify morphological features which were particularly prevalent amongst those nebulae found to host close-binary central stars \cite{miszalski09b}.  This features included jets, rings and filamentary structures, and as such any nebula displaying these structures were considered strong candidates for photometric monitoring in search of variability consistent with central star binarity \cite{miszalski09b}.

\section{A targeted search}

\begin{figure}
\begin{center}
\includegraphics[trim= 100 180 100 200, clip, width=0.6\textwidth]{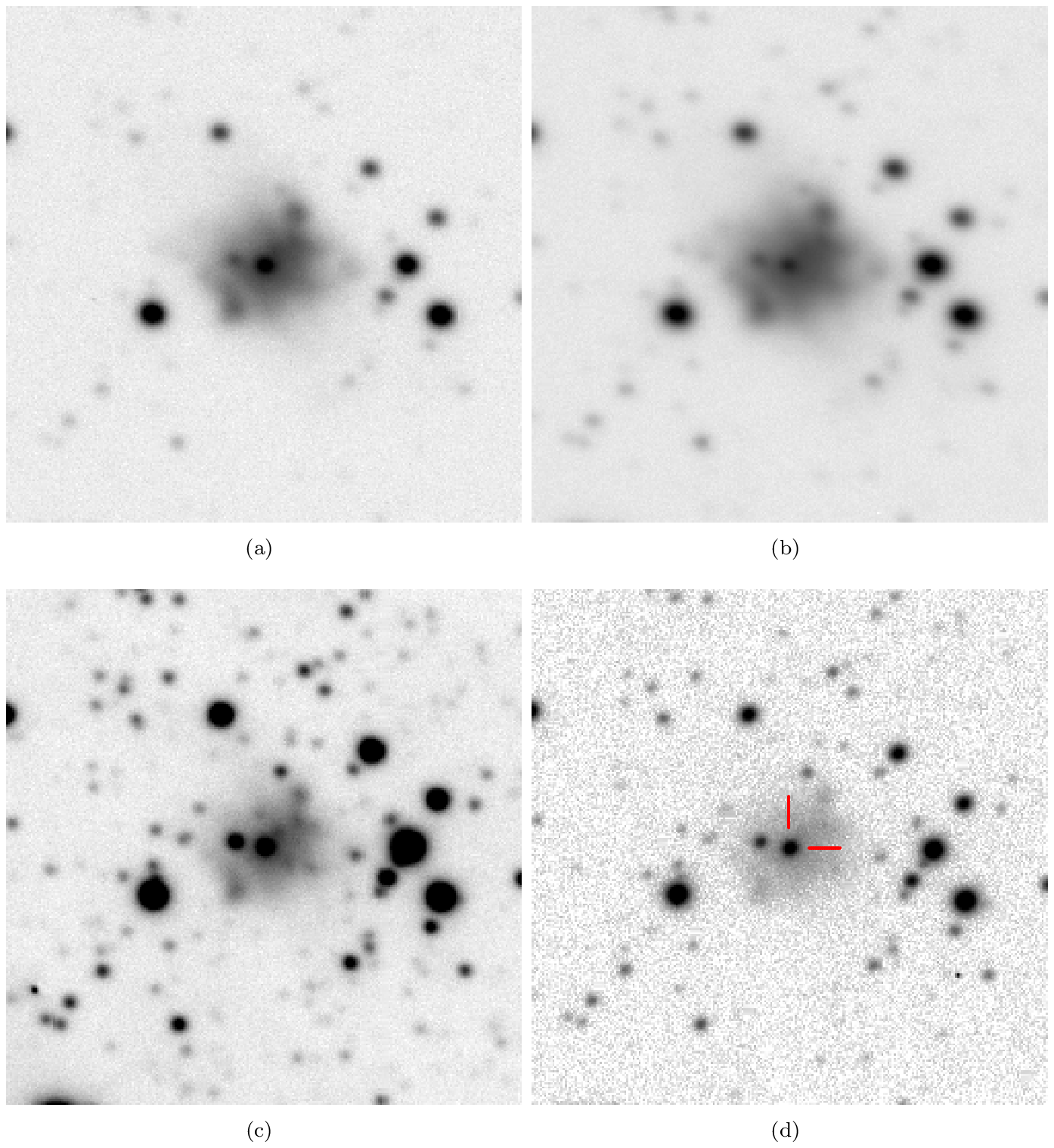}
\end{center}
\caption{\label{filters}Images of the PN Hen 2-155 in (a) B-band, (b) V-band, (c) I-band and (d) H-$\beta$ continuum, demonstrating the value of the use of narrowband, off-emission-line filters in minimising nebular contamination \cite{jones15}.  The location of the central star is marked in the H-$\beta$ continuum image (d).}
\end{figure}

Based on the morphological features identified by \cite{miszalski09b} as being dominant amongst PNe found to host binary central stars, a campaign of photometric monitoring focusing on nebulae displaying those structures was initiated by the author and collaborators.  Many of the best candidate nebulae were bright and irregular presenting obvious problems for photometric monitoring of the central star, where contamination from the nebular background can dilute any observed variability or, worse, introduce spurious variability \cite{miszalski11}. The nebular contamination can often be minimised by careful choice of filter in order to avoid those whose bandpass includes the brightest nebular emission lines (e.g. avoiding R-band which includes the H-$\alpha$ emission line), in many cases the I-band was found to present sufficiently low contamination for successful photometry. However in several cases there was still significant residual nebular emission, for these nebulae it was found that the novel employment of off-emission-line continuum filters reduced the nebula contamination to a minimum acceptable for photometric monitoring.  This approach was successfully employed for several nebulae including, for example, Hen~2-155 \cite{jones15}, where the difference in nebular contamination between broadband filters and the narrowband H-$\beta$ continuum filter is shown in figure \ref{filters}.

Ultimately, the programme of targeted monitoring has been extremely successful, adding more than 10 new binaries to the sample and reaching ever closer to the milestone of 50 known close-binary central stars.  Taking the sample as a whole there are two striking results.  Firstly, the the period distribution is heavily biased towards shorter periods which even when accounting for observational biases does not seem to fit well with population synthesis models \cite{demarco08}.  Secondly, there seems to be an over-abundance of double-degenerate systems (classifying the secondary type based on the shape of the lightcurve).  Double-degenerate central stars should be intrinsically more difficult to detect than those with main sequence companions, and so the fact that roughly 20\% of the known sample of close-binary central stars are double-degenerate has striking consequences for their frequency, in general.

\begin{figure}
\begin{center}
\includegraphics[angle=270,width=0.7\textwidth]{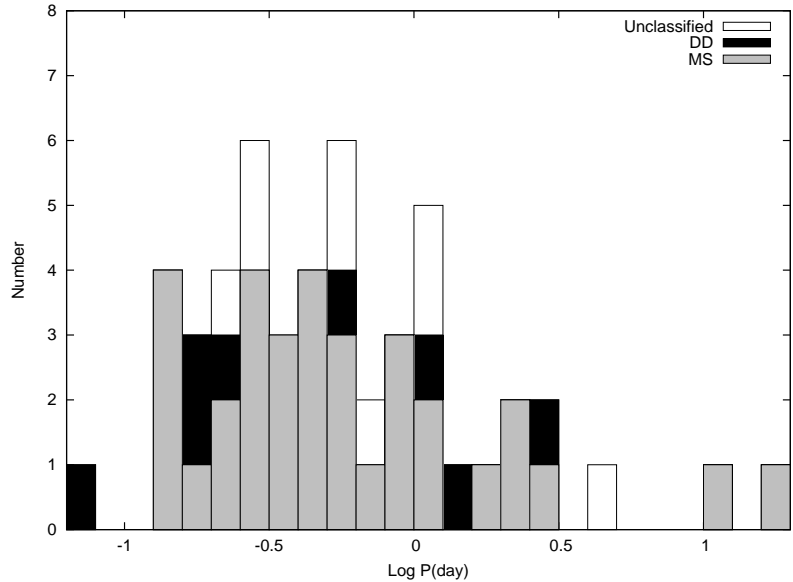}
\end{center}
\caption{\label{periods}The period distribution for all known post-common-envelope binary central stars.  Where known the nature of the secondary is highlighted (main sequence, MS, or degenerate, DD), highlighting the relatively high fraction of double degenerate central binaries.  For full details and references, please see http://drdjones.net/bCSPN}
\end{figure}

Examining individually the newly discovered systems, there are several examples which also have further critical implications for our understanding of common-envelope evolution.  Perhaps the most interesting is the central star system of the Necklace \cite{corradi11}, where the secondary component of the binary was found to be contaminated with AGB material from the primary indicating that significant mass transfer has occurred in the system \cite{miszalski13} - something which is not predicted by current common-envelope models \cite{sandquist98}.

\section{Characterisation}

While some properties of the central binaries can be discerned from light curves alone, full characterisation requires complementary radial velocity studies (with the level of degeneracy in the final parameters also dependent on the presence of eclipses and the single/double-lined nature of the observed radial velocity curve).  These detailed studies are time-consuming and complicated by the presence of nebular emission lines which coincide with many of the central star absorption features which would otherwise be ideal for radial velocity measurements.  As such, only 8 central binary systems have been subjected to detailed modelling based on a combination of photometric and radial velocity observations, with 6 of those being found to host main sequence secondaries which the remaining two are double-degenerate systems \cite{jones15}.  In each of the systems with a main sequence secondary, the secondary is found to be highly inflated with respect to its ZAMS radius, consistent with a period of intense mass transfer prior to entering the common-envelope phase to which the secondary has yet to thermally adjust - again, something which is not predicted by current models.  Of the two double-degenerate systems, one is found to be the best supernova type Ia candidate to-date.  Hen 2-428 has been shown to play host to a pair of degenerate stars, whose total mass exceeds the Chandrasekhar mass, which due to their close orbit will merge in less than a Hubble time \cite{santander-garcia15}.

\section{Conclusions}

Recent works have shown that a common-envelope evolution is responsible for a significant fraction of PNe, with the number of known close-binary central stars now up to approximately 50.  The continued work to search for and characterise close-binary central stars in PNe is critical in understanding not only the formation and evolution of PNe but poorly-understood common-envelope phase, and important phase in the formation of many other astrophysical phenomena.  Even at this early stage, the investigation has revealed many problems with our understanding of the common-envelope phase, perhaps the most interesting being the apparent over-abundance of double-degenerate post-common-envelope systems (particularly interesting in the context of supernova type Ia formation \cite{tsebrenko15}) and the clear evidence of a phase of pre-common-envelope mass transfer, neither of which fit with current theoretical models.  The study of these central stars is also critical in the context of understanding the nature of post-common-envelope PNe and how they relate to the general PN population and other related phenomena \cite{corradi16}.


\medskip


\section*{References}
\begin{thebibliography}{9}
\bibitem{miszalski09a} Miszalski B, Acker A, Moffat A F J, Parker Q A and Udalski A 2009 {\it A\&A} {\bf 496} 813
\bibitem{vanwinckel14} Van Winckel H, Jorissen A, Exter K, Raskin G, Prins S, Perez Padilla J, Merges F and Pessemier W 2014 {\it A\&A} {\bf 563} 10
\bibitem{jones15b} Jones D {\it EAS Publications Series} {\bf 71-72} 113
\bibitem{miszalski09b} Miszalski B, Acker A, Parker Q A and Moffat A F J 2009 {\it A\&A} {\bf 505} 249
\bibitem{miszalski11} Miszalski B, Jones D, Rodr\'iguez-Gil P, Boffin H M J, Corradi R L M and Santander-Garc\'ia M 2011 {\it A\&A} {\bf 531} 158
\bibitem{jones15} Jones D, Boffin H M J, Rodr\'iguez-Gil P, Wesson R, Corradi R L M, Miszalski B and Mohamed S 2015 {\it A\&A} {\bf 580} 19
\bibitem{demarco08} De Marco O, Hillwig T C and Smith A J 2008 {\it AJ} {\bf 136} 323
\bibitem{corradi11} Corradi R L M, Sabin L, Miszalski B, Rodr\'iguez-Gil P, Santander-Garc\'ia M, Jones D, Drew J E, Mampaso A, Barlow M J, Rubio-D\'iez M M, Casares J, Viironen K, Frew D J, Giammanco C, Greimel R and Sale S E 2011 {\it MNRAS} {\bf 410} 1349
\bibitem{miszalski13} Miszalski B, Boffin H M J and Corradi R L M 2013 {\it MNRAS} {\bf 428} 39 
\bibitem{sandquist98} Sandquist E L, Taam R E, Chen X, Bodenheimer P and Burkert A 1998 {\it ApJ} {\bf 500} 909
\bibitem{santander-garcia15} Santander-Garc\'ia M, Rodr\'iguez-Gil P, Corradi R L M, Jones D, Miszalski B, Boffin H M J, Rubio-D\'iez M M and Kotze M M 2015 {\it Nature} {\bf 519} 63
\bibitem{tsebrenko15} Tsebrenko D and Soker N 2015 {\it MNRAS} {\bf 447} 2568
\bibitem{corradi16} Corradi R L M 2016 {\it 11th Pacific Rim Conf. on Stellar Astrophysics}
\end{thebibliography}
\end{document}